\documentclass[10pt,conference]{IEEEtran}
\usepackage[cmex10]{amsmath}

\newtheorem{definition}{Definition} 
\newtheorem{proposition}{Proposition}

\newcommand{\pnt}[1]{{\mbox{\boldmath $#1$}}}

\newcommand{\V}[1]{\mbox{$\mathit{Vars}(#1)$}}

\newcommand{\s}[1]{\mbox{$\{#1\}$}}

\newcommand{\nGz}[2]{$G_{non-\{z\}}$}

\newcommand{\prr}[1]{\mi{Prev}(\boldsymbol{q})}

\newcommand{\mi}[1]{\mathit{#1}}
\newcommand{\ti}[1]{\textit{#1}}
\newcommand{\tb}[1]{\textbf{#1}}

\newcommand{\ttt}{\>\>\>}

\newcommand{\Tt}{\>\>}
\newcommand{\Ss}[1]{\scriptsize{#1}}
\newcommand{\prob}[2]{\mbox{$\exists{#1} [#2]$}}

\newcommand{\Comment}[1]{}

\newcommand{\bm}[1]{{\boldmath $#1$}}

\newcommand{\TS}{\ti{TapSeq}}

\newcommand{\Ro}[1]{\mbox{$R^o(\pnt{#1})$}} 
\newcommand{\Po}[1]{\mbox{$P^o(\pnt{#1})$}} 
\newcommand{\Eo}[1]{\mbox{$E^o(\pnt{#1})$}} 
\newcommand{\inp}[1]{\mbox{$\mi{inp}(\pnt{#1})$}}
\newcommand{\Inp}[1]{\mbox{$\mi{inp}(#1)$}}
\newcommand{\All}{\mbox{$\mi{All\_states}$}}
\newcommand{\Acts}{\mbox{$\mi{Act\_states}$}}
\newcommand{\Curr}{\mbox{$\mi{Curr\_state}$}}
\newcommand{\Poo}[1]{\mbox{$P^o(#1)$}} 
\newcommand{\Rnd}{\mbox{$\mi{RandAlg}$}}

\usepackage{wrapfig}
\usepackage{graphicx}

\begin{document}

\title{Verification of Sequential Circuits by Tests-As-Proofs Paradigm}

\author{\IEEEauthorblockN{Eugene Goldberg, Mitesh Jain,  Panagiotis Manolios} 
\IEEEauthorblockA{
Northeastern University, USA, 
 \{eigold,jmitesh,pete\}@ccs.neu.edu}}

\maketitle

\begin{abstract}
We introduce an algorithm for detection of bugs in sequential circuits.
This algorithm is incomplete i.e. its failure to find a bug breaking a property $P$  does not
imply that  $P$ holds. The appeal of incomplete algorithms
is that they scale better than their complete counterparts. However, to make an incomplete algorithm effective
one needs to guarantee that  the probability of finding a bug is reasonably high. We try to achieve such effectiveness 
by employing the Test-As-Proofs (TAP) paradigm. In our TAP based approach, a counterexample is built as a sequence
of states extracted from  proofs that some   local variations of property $P$  hold. This increases the probability that
  a) a representative set of states is examined and that b)
the considered states are relevant to property $P$.
 We describe an algorithm of test generation based on the TAP paradigm and give preliminary experimental results.
\end{abstract}

\section{Introduction}
Formal  methods have lately made impressive progress in verification of sequential circuits. However,
these methods still do not scale well  enough  to handle large designs. So the  development
of more scalable approaches to sequential verification is an important research direction. One of such approaches is verification 
by simulation i.e. by applying a set of tests.  Simulation is incomplete, which makes it more scalable than formal verification.
An obvious downside of simulation though is that it is limited to  bug hunting.

To make simulation effective it is crucial to increase the probability that, given a buggy circuit, the  part of the search space 
explored by simulation contains a bug. In the case of sequential verification, making simulation effective is especially
challenging for the following reason. Let $P$ be a property of a sequential circuit $\Phi$  to be tested. Suppose that $\Phi$
is buggy. So there is a sequence $\pnt{s_1},\ldots,\pnt{s_k}$ of states of $\Phi$ such that
 \pnt{s_{i+1}} is reachable from \pnt{s_i} in one transition,
$i=1,\ldots,k-1$, state \pnt{s_1} is an initial state and \pnt{s_k} falsifies $P$. Suppose that $k$ is the length of the shortest
counterexample breaking property $P$. This means that no matter how one picks states \pnt{s_i}, $i=1,..,k-1$ they all satisfy 
property $P$.  To make simulation efficient one has  to reduce the set of explored states. But to achieve this goal one must answer
the following tough question: \ti{how does one identify the ``promising'' states if every
 state reachable from an initial state in less than k steps satisfies P}?

In this paper, we address the challenge above using the Tests-As-Proofs (TAP) paradigm~\cite{bridging,tap-10}. The essence of TAP is to treat
a set of tests not as a sample of the search space but as an encoding of a proof that the property in question holds.
So, in a sense, the TAP paradigm reformulates the objective of simulation. Instead of sampling the search space to find  a counterexample
breaking a property $P$, a TAP based algorithm  looks for a hole in a proof that $P$ holds.
A straightforward way of using the TAP paradigm is to generate a set of tests until a counterexample breaking $P$ is found or a test set
encoding a proof  that $P$ holds is generated. In general, this method is very inefficient because checking if a test set encodes a proof
that $P$ holds is computationally hard. There are, however, more practical ways to use TAP. For example, to generate tests for checking
if property $P$ holds, one can first prove that a simpler property  derived from $P$  holds and then use the tests
encoding the obtained proof to verify  $P$ itself. 

In this paper, we describe a TAP based algorithm called \TS~meant for generation of tests for sequential circuits. 
 Let \Po{s}~denote the property that every state reachable from state \pnt{s}
in one transition satisfies property $P$ of a sequential circuit $\Phi$.
 (The superscript 'o' stands for 'one'.) \TS~explores only traces   $\pnt{s_1},\ldots,\pnt{s_k}$
where state \pnt{s_i} is extracted from an encoding of a proof that property \Po{s_{i-1}} holds. 
That is \TS~uses local properties \Po{s} for building a counterexample breaking 
the global property $P$.
The idea here is that, on the one hand, these properties are  related to property $P$ and on the other hand, they are much
easier to prove than $P$. Importantly, a set of states encoding a proof that the property \Po{s} holds is typically a very small subset of 
all states reachable from \pnt{s} in one transition.
So, in a sense, instead of achieving effectiveness of testing by finding  ``promising'' states reachable from \pnt{s} in one transition,
\TS~looks for a \ti{representative} subset of states reachable from \pnt{s} in one transition.

This paper  is structured as follows. 
The TAP paradigm is recalled in Section~\ref{sec:tap_paradigm}. 
In Section~\ref{sec:tap_seq}, our algorithm
for generation of tests for sequential circuits is described. Finally, Section~\ref{sec:experiments}
gives some preliminary experimental results.

\section{Test-As-Proofs Paradigm}
\label{sec:tap_paradigm}
This section is structured as follows. In Subsection~\ref{subsec:res_bnd_pnts}, we recall the notions of a resolution
proof and a boundary point~\cite{esspnts,sat09}. The notion of encoding a resolution proof by a set of points~\cite{bridging,tap-10}
is explained in Subsection~\ref{subsec:proof_encoding}. Subsection~\ref{subsec:tap_comb_circ}
recalls the Tests-As-Proofs paradigm~\cite{bridging,tap-10} by the example of testing  combinational circuits.
%
%

\subsection{Resolution and Boundary Points}
\label{subsec:res_bnd_pnts}
%
%
\begin{definition}
A \tb{literal} of a Boolean variable $v$ is $v$ itself (positive literal) or
the negation of $v$ (negative literal). A \tb{clause} $C$ is a disjunction of literals.
We will assume that a clause $C$ cannot have two literals of the same variable.
A \tb{Conjunctive-Normal Form (CNF)} $F$ is a conjunction of clauses.
We will also consider $F$ as just a set of clauses. So, for instance, the CNF formula
$F \wedge G$ can  also be represented as $F \cup G$.
\end{definition}
%
%
\begin{definition}
Let $X$ be a set of Boolean variables. An \tb{assignment} \pnt{q} to variables of $X$
is a mapping $Z \rightarrow \s{0,1}$ where $Z \subseteq X$. 
We will also  consider \pnt{q} as a set of value assignments to the individual variables of $Z$.
If $Z=X$, the assignment \pnt{q}
is called \tb{complete}. We will also refer to a complete assignment as a \tb{point}.

\end{definition}
%
%
\begin{definition}
Let $F$ be a CNF formula and $C$ be a clause.
Denote by {\boldmath $\mi{Vars}(F)$} (respectively {\boldmath $\mi{Vars(C)}$}) the set of variables
of $F$ (respectively $C$). Let \pnt{q} be an assignment. We denote the set of variables assigned
in \pnt{q} as  {\boldmath $\mi{Vars(q)}$}.
\end{definition}
%
%
\begin{definition}
Let $v$ be a Boolean variable. A literal of $v$ is said to be satisfied (falsified) by an assignment to $v$
if it evaluates to 1 (respectively to 0) by this assignment.
A clause $C$ is said to be \tb{satisfied} (respectively \tb{falsified}) by an assignment \pnt{q} if
a literal of $C$ is satisfied by \pnt{q} (respectively all literals of $C$ are falsified by \pnt{q}).
A CNF formula $F$ is \tb{satisfied} (respectively \tb{falsified}) by an assignment \pnt{q} if every 
clause of $F$ is satisfied by \pnt{q} (respectively at least one clause of $F$ is falsified by \pnt{q}).
\end{definition}
%
%
\begin{definition}
Let $C'~\vee~v$ and $C''~\vee~\overline{v}$ be two clauses such that 
no variable of $\V{C'} \cap \V{C''}$ has opposite literals in $C'$ and $C''$. 
The clause $C' \vee C''$ is called the \tb{resolvent} of the \tb{parent clauses} $C'~\vee~v$ and $C''~\vee~\overline{v}$.
This resolvent is said to be obtained  by \tb{resolution} of the parent clauses on $v$. 
Clauses $C'~\vee~v$ and $C''~\vee~\overline{v}$ are called \tb{resolvable} on  $v$.
\end{definition}
%
%
\begin{definition}
Let $F$ be a CNF formula. A clause $C$ is said to be \tb{derived from} \pnt{F} by a set of resolutions
$r_1,\ldots,r_k$ if
\begin{itemize}
\item the resolvent of resolution $r_k$ is clause $C$,
\item the parent clauses of  resolution $r_i$, $i=1,\ldots,k$ are either clauses
of $F$ or resolvents of resolutions $r_j$ where $j < i$.
\end{itemize}
We will call the sequence $r_1,\ldots,r_k$ a \tb{resolution derivation} of clause $C$ from $F$.
\end{definition}
%
%
\begin{proposition}
The resolution proof system based on the operation of resolution 
is complete in the following sense. Given a CNF formula $F$ and a clause $C$ such
that $F \rightarrow C$, there is a resolution derivation of clause  $C'$ from $F$ such that
$C' \rightarrow C$. In particular, if $F$ is unsatisfiable,  one can always derive
 an \tb{empty clause} from $F$ i.e. a clause that has  no literals and so cannot be
satisfied. Derivation of an empty clause from $F$ is called a \tb{resolution proof} that
$F$ is unsatisfiable. 
\end{proposition}
%
%
\begin{definition}
Let $F$ be a CNF formula and \pnt{p} be a complete assignment
to  \V{F}. Point \pnt{p} is called a \pnt{v}\tb{-boundary point}
of $F$ if
\begin{itemize}
\item \pnt{p} falsifies $F$,
\item  every clause of $F$ falsified by \pnt{p} has variable $v$.
\end{itemize}
\end{definition}
Proposition~\ref{prop:bnd_pnts} below shows that boundary points characterize  ``mandatory''
fragments of resolution proofs.
%
%
\begin{proposition}
\label{prop:bnd_pnts}
Let $F$ be an unsatisfiable formula and \pnt{p} be a $v$-boundary point of $F$.
Then any resolution proof that $F$ is unsatisfiable contains a resolution $r$ such that
\begin{itemize}
\item $r$ is a resolution on variable $v$,
\item the resolvent produced by $r$ is falsified by \pnt{p}.
\end{itemize}
\end{proposition}

%
%
\subsection{Set of Points Encoding a Resolution Proof}
\label{subsec:proof_encoding}
%
%
\begin{definition}
\label{def:legal_res}
Let $X$ be a set of Boolean variables and $P$ be a set of points i.e. complete assignments 
to $X$. Let $C'$ and $C''$ be two clauses such that
\begin{itemize}
\item $(\V{C'} \cup \V{C''}) \subseteq X$,
\item $C'$ and $C''$ are resolvable on variable $v$.
\end{itemize}
Resolving  $C'$ and $C''$ on $v$ is said to be \tb{legal} with respect to $P$ 
if there are points $\pnt{p'},\pnt{p''} \in P$ such that
\begin{itemize}
\item \pnt{p'} falsifies $C'$ and \pnt{p''} falsifies $C''$,
\item \pnt{p'} and \pnt{p''} are different only in the value of $v$.
\end{itemize}
\end{definition}
%
%
\begin{proposition}
\label{prop:legal_via_resolvent}
Let clause $C$ be obtained by resolving clauses $C'$ and $C''$ on variable $v$.
Then points \pnt{p'} and \pnt{p''} make this resolution legal iff both \pnt{p'} and \pnt{p''}
falsify $C$ and are different only in variable $v$.
\end{proposition}
%
%
\begin{definition}
Let $F$ be an unsatisfiable CNF formula and $P$ be a set of complete assignments to \V{F}.
Suppose, there is a  resolution proof  $R=r_1,\ldots,r_k$ that $F$ is unsatisfiable
 such that every resolution $r_i,i=1,\ldots,k$ is legal with respect to $P$.  We will say then that the set of points $P$ 
\tb{encodes proof} \pnt{R}. More generally, we will say that a set of points $P$ encodes an unspecified
 resolution proof that $F$ is unsatisfiable
if there is a resolution proof of unsatisfiability of $F$ encoded by $P$.
\end{definition}

There is  a simple but very inefficient 
procedure~\cite{tap-10} for checking  if a set of points $P$ encodes a resolution proof that a CNF formula $F$ is unsatisfiable.
This procedure starts by making sure  that every point of $P$ falsifies $F$. If not, then $F$ is satisfiable.
Otherwise, all resolution operations that are legal with respect to set of points $P$ are performed. If an empty clause is derived
then $P$ encodes a proof that $F$ is unsatisfiable. Otherwise, $P$ is too small and needs to be expanded to either include
an assignment satisfying $F$ or to encode a proof that $F$ is unsatisfiable.

Obviously, the procedure above is impractical. Unfortunately, no efficient procedure for checking if a set of points
encodes  a resolution proof is known. On the contrary, the \ti{reverse} procedure of finding
a set $P$ encoding a given resolution proof $r_1,\ldots,r_k$ is trivial. The idea of this procedure is to start with an empty
set of points $P$ and then add points that makes resolutions of the proof legal. Let $r_i$ be a resolution in which
clauses $C'$ and $C''$ are resolved on variable $v$ producing resolvent $C$. From Proposition~\ref{prop:legal_via_resolvent} it follows
that  to make $r_i$ legal one just needs to add to $P$ points \pnt{p'} and
\pnt{p''} that falsify $C$ and are different only in value of $v$. So the upper bound on the size of $P$ is $2*k$ because one needs
two points per resolution. In reality, the size of $P$ may be much smaller because two-point sets legalizing 
different resolutions $r_i$ and $r_j$ may overlap.

%
%
\subsection{Test-as-Proofs Paradigm}
\label{subsec:tap_comb_circ}
In this subsection, we introduce the Tests-As-Proofs (TAP) paradigm by showing how one can use tests to encode
a proof of a property of a combinational circuit.  Let $N(X,Y,z)$ be a single-output combinational circuit.
Here $X$ and $Y$ denote input and internal variables of $N$ respectively and $z$ denotes the output of $N$.
We will assume that the fact the $N$ evaluates only to 0 means that a combinational property holds. (For instance,
$N$ can be the miter of two combinational circuits $M'$, $M''$ checked for equivalence. Then the fact that $N$ always evaluates to 0
means that $M'$ and $M''$ are functionally equivalent.) If $N$ evaluates to 1 for some input
assignment \pnt{x}, then property specified by $N$ does not hold and \pnt{x} is a counterexample.

Let $F_N(X,Y,z)$ be a CNF formula specifying circuit $N$, i.e. a satisfying assignment of $F_N$ corresponds
to a consistent assignment to gates of $N$ and vice versa. Let $F$ denote the formula $F_N \wedge z$. The
satisfiability of $F$ means that, for some input assignment, $N$ evaluates to 1 and so there is a bug.

Suppose that
$F$ is unsatisfiable and   $\Psi = \s{r_1,\ldots,r_k}$ is a resolution proof of that. Let \pnt{p} be a complete
assignment to \V{F}. Denote by \inp{p} be the projection of \pnt{p} onto the set of input variables $X$. 
 Let $E=\s{\pnt{p_1},\ldots,\pnt{p_m}}$ be a set of points encoding $\Psi$. Let \Inp{E} denote
$E=\s{\inp{p_1},\ldots,\inp{p_m}}$. Notice that \inp{p_i} may be equal to \inp{p_j} for two different points
\pnt{p_i},\pnt{p_j} of $E$. We will assume that \Inp{E} does not contain duplicates. We will say
that the \tb{set of tests}  $T=\s{\pnt{x_1},\ldots,\pnt{x_d}}$  \tb{encodes proof} $\Psi$  
if there is a set of points $E$ encoding $\Psi$ such that $T = \Inp{E}$. Similarly, set $T$ encodes an unspecified
resolution proof if there is a set of points $E$ encoding a resolution proof such that $T = \Inp{E}$.

As we mentioned in Subsection~\ref{subsec:proof_encoding}, the size of a set of points $E$
 encoding a proof $\Psi$ is bounded by $2*|\Psi|$ where $|\Psi|$ is
the number of resolutions in $\Psi$. Since $|\Inp{E}| \leq |E|$, the same applies to the size of a set of tests
encoding $\Psi$. In reality, as we mentioned above, $|\Inp{E}|$ may be drastically smaller than $|E|$ because
different points of $E$ may have identical projections onto the set of input variables.

The relation between tests and proofs implies that testing can be viewed as finding an encoding
of a proof that the property in question holds rather than sampling the search space. We will
refer to such a point of view at the \tb{Tests-As-Proofs (TAP) paradigm}. There are numerous
ways to use the TAP paradigm in practice. One of them is to build a test set encoding a proof
that a property of a circuit holds and apply it in a different situation. (For instance, this set
of tests can be used to check if this circuit still has the same property after a modification.)

In Subsection~\ref{subsec:proof_encoding}, we outlined a trivial procedure of building
a set of points $E$ encoding a known proof $\Psi$ that $F$ is unsatisfiable. However, this procedure cannot  guarantee 
that the set of tests \Inp{E} extracted from $E$ has high quality. To produce a test set
of high-quality one needs to extract them from a set of points $E$ forming
a \tb{tight encoding} of $\Psi$. The intuition here is that the closer  a set of points $E$
encoding $\Psi$ to $\Psi$, the higher the quality of tests \Inp{E}. By proximity of $E$ to $\Psi$ we mean 
that $E$ makes legal the smallest possible set of resolutions that are not in $\Psi$.

 Informally, building a tight proof encoding means  that when looking for points \pnt{p'},\pnt{p''} legalizing
resolution of clauses $C'$ and $C''$ one needs to make \pnt{p'},\pnt{p''} satisfy as many clauses of $F$ as possible. (In particular,
if a clause $C$ of $F$ is satisfied by every point of $E$, then $C$ is redundant in a
proof encoded by $E$. This is because any resolution involving $C$ is illegal with respect to $E$.) One way to build a tight proof encoding is
to require that \pnt{p'},\pnt{p''} are $v$-boundary points of $F$ where $v$ is the variable on which $C'$ and $C''$
are resolved. The high quality of tests extracted from boundary points has been confirmed in~\cite{tap-10}.

\section{TAP Based Generation Of Tests For Sequential Circuits}
\label{sec:tap_seq}
In this section, we describe an  algorithm based on the TAP paradigm  meant for testing sequential circuits. 
We will refer to this algorithm as \TS. This section is structured as follows.
In Subsection~\ref{subsec:definitions}, some basic definitions of sequential verification are listed.
A high-level view of \TS~is given in Subsection~\ref{subsec:high_level_view}. Subsection~\ref{subsec:detailed_descr}
describes \TS~in more detail.
%
%
\subsection{Some Definitions}
\label{subsec:definitions}
%
%
\begin{definition}
\label{def:trans_relation}
A sequential circuit $\Phi$ is specified by a pair of predicates
 $(I,T)$ over Boolean variables.
Here $T(S,S',Z)$ is  the \tb{transition relation} of $\Phi$ where  $S,S'$ are the sets of
 present and next state variables respectively, and  $Z$ is the set of combinational variables.
 Predicate $I(S)$ specifies the set of initial states of $\Phi$. 
 We will denote the \tb{input variables} of $\Phi$ by $X$ where $X \subseteq Z$.
\end{definition}

%
%
\begin{definition}
Let  pair $(I(S)$,$T(S,S',Z))$ specify a circuit $\Phi$.
A complete assignment \pnt{s} to variables of $S$ (respectively $S'$) is called
 \tb{a state} (respectively \tb{next state})  of $\Phi$.
\end{definition}

%
%
\begin{definition}
Let $\Phi$ be a circuit specified by pair $(I,T)$.
 A sequence of states  $\pnt{s_1},\ldots,\pnt{s_k}$ is called a \tb{trace} 
if $I(\pnt{s_1}) = 1$ and  \prob{Z}{T(\pnt{s_i},\pnt{s_{i+1}},Z)}=1 for every $i$ where 
$1 \leq i \leq k-1$.
\end{definition}
%
%
\begin{definition}
Let $\Phi$ be a circuit specified by pair $(I,T)$.
The state \pnt{s} is called \tb{reachable} by $\Phi$ if there is a trace
ending in state \pnt{s}. 
Denote by \bm{R(S)} a predicate specifying the set of \tb{all reachable states} 
of $\Phi$. That is
$R(\pnt{s})=1$ if and only if state \pnt{s} is reachable.
\end{definition}
%
%
\begin{definition}
In this paper, we consider the problem of property checking. Let $\Phi$ be a circuit
 specified by pair $(I,T)$.
A property of $\Phi$ is specified by a predicate $P(S)$ describing the set of states
 where this property holds (i.e. the set of \tb{good states}). So the predicate
$\overline{P}$ specifies the set of \tb{bad states}.
For the sake of simplicity, we will refer to the property specified by $P$ as 
\tb{property} \bm{P}.
 We will say that  property $P$ holds for $\Phi$ if $R \wedge \overline{P} \equiv 0$. 
\end{definition}
%
%
\begin{definition}
Let $\Phi$ be a circuit specified by pair $(I,T)$. Let $P$ be a property
of $\Phi$ and \pnt{s} be a state of $\Phi$. Denote by {\boldmath $R^o(s)$} the set
of all states of $\Phi$ that are reachable from \pnt{s} in one transition.
Denote by {\boldmath $P^o(s)$} the property that holds iff the property $P$
holds for every state of \Ro{s}.
\end{definition}

%
%
\subsection{High-level View of \TS}
\label{subsec:high_level_view}
Let $\Phi$ be a sequential circuit specified by pair $(I,T)$. Let $P$ be a property of $\Phi$ to be verified.
The pseudocode of \TS~is given in Figure~\ref{fig:tap_seq}.
\TS~is incomplete i.e. it can  build a counterexample breaking $P$ but cannot prove that $P$ holds.
 For the sake of simplicity we will assume  that there is only
one state \pnt{s_1} satisfying $I$ i.e. $\Phi$ has only one initial state.

First, \TS~checks if  property $P^o(\pnt{s_1})$ 
 holds. If not, then there is a bad state $\pnt{s_2} \in \Ro{s_1}$  and  \pnt{s_1},\pnt{s_2} form a counterexample.
Otherwise, a resolution proof is generated stating that \Po{s_1} holds and a set of states \Eo{s_1} is extracted
from an encoding of this proof.
Here \Eo{s_1} is a subset of  \Ro{s_1}. Then the same
procedure repeats for the states of \Ro{s_1}. That  is for every state $\pnt{s} \in \Ro{s_1}$, \TS~checks the property
\Po{s}.  If it does  not hold,
then a state $\pnt{s^*}\in \Ro{s}$ breaks $P$  and \pnt{s_1},\pnt{s},\pnt{s^*} form a counterexample.  Otherwise, new states \Eo{s}
are extracted from an encoding of  a proof that \Ro{s} holds.  

\TS~maintains the set  \ti{All\_states} of all visited states.  This allows one to avoid visiting the same state more than once.
\TS~terminates in two cases.
\begin{itemize}
\item A bad state is reached (property $P$ does not hold).
\item No new states are extracted from encodings of  proofs of properties \Po{\pnt{s}}, $\pnt{s} \in \mi{All\_states}$.
      In this case, we will say that \TS~reached a \tb{convergence point}.
\end{itemize}

%
%
\subsection{More Detailed Description of \TS}
\label{subsec:detailed_descr}

%
%
\setlength{\intextsep}{10pt}
\setlength{\textfloatsep}{10pt}
\begin{figure}
\small
\vspace{-10pt}
\begin{tabbing}
aaa\=bb\=cc\= dd\=eeeeeeeee \= \kill
// \TS~returns \ti{bug} if a reachable bad state is found\\
// Otherwise \TS~returns \ti{no\_bug\_found}  \\
// \\
\TS$(I,T,P)$\{\\
\tb{\scriptsize{1}}\> if ($I \wedge \overline{P} \not\equiv \emptyset$) return($\mi{bug}$); \\
\tb{\scriptsize{2}}\> $\mi{All\_states} := \{\mi{init\_state}(I)\}$; \\
\tb{\scriptsize{3}}\> $\mi{Act\_states} := \All$; \\ 
\tb{\scriptsize{4}}\> while ($\mi{Act\_states} \neq \emptyset$)\{ \\
\tb{\scriptsize{5}}\Tt  $\mi{Curr\_state} := \mi{pick\_state}(\mi{Act\_states})$; \\
\tb{\scriptsize{6}}\Tt  $\mi{Act\_states} := \mi{Act\_states} \setminus \{\mi{Curr\_state}\}$; \\
\tb{\scriptsize{7}}\Tt $\mi{sat} := \mi{enc\_proof}(\All,\Acts,$\\
                   \>\>\>\>\> $\mi{Curr\_state},T,P)$; \\
\tb{\scriptsize{8}}\Tt if ($\mi{sat}$) return($\mi{bug}$); \}\\
\tb{\scriptsize{9}}\> return($\mi{no\_bug\_found})$;\}\\

\end{tabbing} 
\vspace{-10pt}
\caption{\ti{Pseudocode of \TS}}
\label{fig:tap_seq}
\end{figure}

\TS~starts by checking if the initial state breaks property $P$ (line 1 of Figure~\ref{fig:tap_seq}).
 If it does, then \TS~terminates reporting a bug. Otherwise,
variables \All~and \Acts~are initialized with the initial state.  As we mentioned above, \All~specifies the set of all visited states.
\Acts~is a subset of \All. A state \pnt{s} remains in \Acts~until  the validity of property \Po{s} is established.

%
%
\setlength{\intextsep}{10pt}
\setlength{\textfloatsep}{10pt}
\begin{figure}[htb]
\small
\vspace{10pt}
\begin{tabbing}
aaa\=bb\=cc\= dd \= eeeeee \= \kill
$\mi{enc\_proof}(\All,\Acts,\mi{Curr\_state},T,P)$\{\\
\tb{\scriptsize{1}}\> $F = \mi{cnf}(\mi{Curr\_state})  \wedge T \wedge \overline{P'}$;\\
\tb{\scriptsize{2}}\> $(\Psi,\mi{sat}) := \mi{gen\_proof}(F)$;\\ 
\tb{\scriptsize{3}}\> if ($\mi{sat}$) return($\mi{true}$);\\
\tb{\scriptsize{4}}\>  $\mi{enc\_resol}(\All,\Acts,\Psi,F)$;\\
\tb{\scriptsize{5}}\> return($\mi{false}$); \}\\
\end{tabbing} 
\vspace{-10pt}
\caption{\ti{Pseudocode of enc\_proof}}
\label{fig:enc_frw_proof}
\end{figure}

The main work is done by \TS~in a 'while' loop (lines 4-8). First, \TS~picks a  state from \Acts~ and 
removes the former from the latter. This state is assigned to variable \Curr~that is used to specify the state currently processed by \TS.
Notice that every state assigned to \Curr~is reachable from the initial state.
Then \TS~checks if property \Poo{\Curr} holds (line 7). If not, then \TS~reports the presence of a bug.
Otherwise, a proof that \Poo{\Curr} holds is generated. This proof is encoded and  new states (if any) are added to
\All~and~\Acts~by procedure \ti{enc\_proof}. Then a new iteration begins. Iterations go on as long as \Acts~is not empty.
Once a convergence point is reached (i.e. \Acts~becomes empty), \TS~terminates reporting that no bug was found.

The pseudocode of the \ti{enc\_proof} procedure is shown in Figure~\ref{fig:enc_frw_proof}. First, a CNF formula $F$ is formed (line 1) that
is satisfiable iff property \Poo{\Curr} does not hold. The satisfiability of $F$ is checked in line 2. If $F$ is satisfiable, then 
\ti{enc\_proof} terminates (line 3). Otherwise, a proof $\Psi$ of unsatisfiability of $F$ is generated. Resolutions of $\Psi$ 
are encoded by \ti{enc\_resol} procedure shown in Figure~\ref{fig:enc_resol}.

%
%
\setlength{\intextsep}{10pt}
\setlength{\textfloatsep}{10pt}
\begin{figure}
\small
\begin{tabbing}
aaa\=bb\=cc\= dd \= eeeeee \= \kill
$\mi{enc\_resol}(\All,\Acts,\Psi,F,\mi{Curr\_state},T)$\{\\
\tb{\scriptsize{1}}\> while ($\Psi \neq \emptyset$) \{\\
\tb{\scriptsize{2}}\Tt $(C,v):= \mi{extract\_resolution}(\Psi)$\\ 
\tb{\scriptsize{3}}\Tt $\Psi := \Psi \setminus \s{(C,v)}$; \\
\tb{\scriptsize{4}}\Tt $\pnt{p} := \mi{enc\_clause}(F,C,v,\mi{Curr\_state})$; \\
\tb{\scriptsize{5}}\Tt if ($\pnt{p} = \mi{nil}$) continue; \\
\tb{\scriptsize{6}}\Tt $\mi{update\_states}(\mi{All\_states},\pnt{p},T)$; \}\}\\
\end{tabbing} 
\vspace{-10pt}
\caption{\ti{Pseudocode of} $\mi{encode\_resol}$}
\label{fig:enc_resol}
\end{figure}

Procedure \ti{enc\_resol} loops over resolutions of proof $\Psi$. First, it extracts a new resolution ($C,v$) of $\Psi$ and removes it from the latter.
Here $C$ is the resolvent  and $v$ is the variable on which the parent clauses of $C$ were resolved. Then, a $v$-boundary point \pnt{p}
of $F$ falsifying $C$ is generated by procedure \ti{enc\_clause}.  
From Proposition~\ref{prop:legal_via_resolvent} it follows, that \pnt{p} and the point obtained from \pnt{p} by flipping the value of $v$
legalize the resolution specified by $C$ and $v$. We want \pnt{p} to be a $v$-boundary point to make our proof encoding \ti{tight}.
If \pnt{p} does not exist, \ti{enc\_resolutions} starts a new iteration. Otherwise, procedure \ti{update\_states} is called
to update sets \All~and \Acts.

%
%
\setlength{\intextsep}{10pt}
\setlength{\textfloatsep}{10pt}
\begin{figure}
\small
\vspace{10pt}
\begin{tabbing}
aaa\=bb\=cc\= dd \= eeeeee \= \kill
$\mi{enc\_clause}(F,C,v,\mi{Curr\_state})$\{\\
\tb{\scriptsize{1}}\> $\pnt{p} := \mi{find\_sat\_assgn}((F \cup  \overline{C}) \setminus F^{\{v\}})$;\\ 
\tb{\scriptsize{2}}\> if ($\pnt{p} = \mi{nil}$) return($\mi{nil}$);\\
\tb{\scriptsize{3}}\> $\pnt{p} := \mi{assign\_var}(\pnt{p},v,Curr\_state)$;\\
\tb{\scriptsize{4}}\> return(\pnt{p}); \}\\
\end{tabbing} 
\vspace{-10pt}
\caption{\ti{Pseudocode of enc\_clause}}
\label{fig:enc_clause}
\end{figure}

The pseudocode of procedure \ti{enc\_clause} is shown in Figure~\ref{fig:enc_clause}. This procedure computes a $v$-boundary
point of formula $F$ that falsifies a resolvent clause $C$. This is done by finding an assignment satisfying formula 
 $F \cup \overline{C} \setminus F^v$ where $F^v$ is the set of clauses of $F$ containing variable $v$. Notice that if \pnt{p}
satisfies  $F \cup \overline{C} \setminus F^v$ then it satisfies all the clauses of $F$ but some clauses containing variable $v$. In
other words, \pnt{p} is a $v$-boundary point of $F$.  After computing \pnt{p}, the value of variable $v$ is set in \pnt{p} (line 3).
If $v \not\in S$, then the value of $v$ is set arbitrarily. Otherwise, $v$ is assigned the same value as in \Curr. This is
done to guarantee that the new states generated by  \ti{update\_states} are reachable from \Curr~in one transition.

The fact that one uses only $v$-boundary points that agree with the values of \Curr~means that  proof $\Psi$ is encoded only partially.
Namely, this encoding does not  legalize resolutions on variables of $S$. This is done to simplify \TS. We are going to fix 
this problem in future versions of \TS.

%
%
\setlength{\intextsep}{10pt}
\setlength{\textfloatsep}{10pt}
\begin{figure}
\small
\vspace{10pt}
\begin{tabbing}
aaa\=bb\=cc\= dd \= eeeeee \= \kill
$\mi{update\_states}(\mi{All\_states},\pnt{p},T)$\{\\
\tb{\scriptsize{1}}\>(\pnt{s},\pnt{x}) := $\mi{extract\_state\_input}$(\pnt{p}); \\
\tb{\scriptsize{2}}\> $\pnt{s^*} := \ti{find\_next\_state}(\pnt{s},\pnt{x},T)$; \\ 
\tb{\scriptsize{3}}\> if ($\pnt{s^*} \in \mi{All\_states}$) return; \\
\tb{\scriptsize{4}}\> $\mi{All\_states} := \mi{All\_states} \cup \s{\pnt{s^*}}$; \\
\tb{\scriptsize{5}}\> $\mi{Act\_states} := \mi{Act\_states} \cup \s{\pnt{s^*}}$; \}\\
\end{tabbing} 
\vspace{-10pt}
\caption{\ti{Pseudocode of update\_states}}
\label{fig:update_states}
\end{figure}

Figure~\ref{fig:update_states} shows the pseudocode of procedure \ti{update\_states}.
First, the assignments (\pnt{s},\pnt{x}) to variables of $S$ and $X$ (i.e. present state and input variables) are extracted
from a $v$-boundary point found by procedure \ti{enc\_clause}. Then the transition relation $T$ is used to compute
the state \pnt{s^*}  to which circuit $\Phi$ switches from state  \pnt{s} under the input assignment \pnt{x}.
If \pnt{s^*} is a new state, it is added to \All~and \Acts.

\section{Experimental Results}
\label{sec:experiments}

In this  section, we describe two experiments  conducted to  evaluate the performance of \TS.
This section is structured as follows. In Subsection~\ref{subsec:rand_alg}, we describe an
 algorithm of random test generation that we compared with \TS. Some details of the implementation of \TS~we used
in experiments are given in Subsection~\ref{subsec:impl_tap_seq}. The first and second experiments
are described in Subsections~\ref{subsec:first_exper} and~\ref{subsec:second_exper} respectively.
%
%
\subsection{Random Algorithm We Used in Experiments}
\label{subsec:rand_alg}

%
%
\setlength{\intextsep}{10pt}
\setlength{\textfloatsep}{10pt}
\begin{figure}
\small
\begin{tabbing}
aaa\=bb\=cc\= dd \= eeeeee \= \kill
\Rnd$(I,T,P,\mi{max\_tries},\mi{max\_length})$\{\\
\tb{\scriptsize{1}}\> if ($I \wedge \overline{P} \not\equiv \emptyset$) return($\mi{bug}$); \\
\tb{\scriptsize{2}}\> $\mi{Curr\_state} :=  \mi{set\_init\_state}(I)$; \\
\tb{\scriptsize{3}} \> $\mi{length} := 0$; $\mi{tries} := 0$; \\
\tb{\scriptsize{4}} \> while ($\mi{tries} \leq \mi{max\_tries}$) \{\\
\tb{\scriptsize{5}} \Tt if ($\mi{length} > \mi{max\_length}$) \{ \\
\tb{\scriptsize{6}} \ttt  $\mi{length} := 0$; $\mi{tries}\scriptstyle{++}$; \\
\tb{\scriptsize{7}} \ttt $\mi{Curr\_state} := \mi{set\_init\_state}(I)$; \\
\tb{\scriptsize{8}} \ttt continue;\} \\
\tb{\scriptsize{9}}\Tt$F := \mi{cnf}(\mi{Curr\_state}) \wedge T \wedge \overline{P}$;\\
\tb{\scriptsize{10}} \Tt if $(\mi{satisf}(F))$ return(\ti{bug}); \\
\tb{\scriptsize{11}} \Tt $\pnt{x} := \mi{gen\_rand\_input}(X)$; \\
\tb{\scriptsize{12}}\Tt $\mi{Curr\_state} := \ti{next\_state}(T,\pnt{x},\mi{Curr\_state})$; \\
\tb{\scriptsize{13}}\Tt $\mi{length}\scriptstyle{++}$; \} \\
\tb{\scriptsize{14}} \> return(\ti{no\_bug\_found}); \} \\
\end{tabbing} 
\vspace{-10pt}
\caption{\ti{Algorithm for generation of  counterexamples randomly}}
\label{fig:rand_alg}
\end{figure}

In this  subsection, we describe an algorithm of random test generation we used in the first experiment.
We will refer to this algorithm as \Rnd. 
The pseudocode of \Rnd~is shown in Figure~\ref{fig:rand_alg}.
The set of counterexamples generated by \Rnd~is controlled
by parameters \ti{max\_tries} and \ti{max\_length}. The value of \ti{max\_tries} limits the number
of generated  counterexamples while \ti{max\_length} sets the limit to the number of states
in a counterexample. The length of the current counterexample and the number of counterexamples 
generated so far are specified by variables \ti{length} and \ti{tries} respectively.

\Rnd~maintains variable \Curr~specifying a state reachable from the initial state
that is  currently processed by \Rnd. At the beginning, \Curr~is set to the initial state (line 2).
The main work is done in the 'while' loop (lines 4-13). If the value of \ti{length} 
exceeds \ti{max\_length}, a new counterexample is started  and the value
of \ti{tries} is incremented (lines 5-8). Otherwise, \Rnd~checks if \Curr~satisfies property $P$.
If not, then  \Rnd~returns value \ti{bug}.
Otherwise, \Rnd~randomly generates an assignment
\pnt{x} to input variables $X$ (line 11). Then \pnt{x} is used to generate a new state
that is the state to which the circuit switches from state \Curr~under input assignment \pnt{x}
(line 12). After that, the length of the current counterexample is incremented and a new iteration begins.
%
%
\subsection{Implementation of \TS}
\label{subsec:impl_tap_seq}
In the pseudocode of \TS~given in Figure~\ref{fig:tap_seq}, we did not clarify in what order states were extracted
from \Acts~in the 'while' loop. The two extremes are  depth-first and breadth-first orders. The depth-first order is to
 first process the state of \Acts~the is the farthest
from the initial state (in terms of transitions).  On the contrary, the breadth-first order, is to first process the state that is the
closest to the initial state. In the breadth-first variant of \TS, states are processed  one time frame after another. We assume here
that $i$-th time frame consists of the states of \All~that can be reached from the initial state in $i$ transitions.
That is, in the breadth-first variant, a state of \Acts~of $i$-th time frame is processed only after every state
of every $j$-th time frame where $j < i$ has been processed and removed from \Acts.
Obviously, by imposing a particular order of extracting states from \Acts~one can also have modifications of \TS~ that are different
from the two extremes above. \ti{In this paper, we report results of  a breadth-first implementation of \TS}.

In the experiments, we ran two versions of \TS: randomized and non-randomized. The difference between these versions is in finding
boundary points used to encode proofs. In the randomized version, the internal SAT-solver called  to find boundary points
had some randomization in its decision making. Namely, the phase of every 10-th decision assignment was chosen randomly. 
The reason for such randomization is explained in Subsection~\ref{subsec:first_exper}.
%
%
\subsection{First Experiment: Comparison of \TS~with \Rnd}
\label{subsec:first_exper}
The objective of the first experiment was to compare \TS~with \Rnd.
In this comparison we used 314 buggy benchmarks of the HWMCC-10 competition.
 78 benchmarks of this set were trivial: the initial state did not satisfy the property to be verified. 
We excluded them from consideration.
The results of the  experiment on non-trivial benchmarks are summarized in Table~\ref{tbl:rand_vers_tap}.

%
%
\setlength{\abovecaptionskip}{2pt}   
\setlength{\belowcaptionskip}{5pt}   
\setlength{\intextsep}{1pt}
\setlength{\floatsep}{1pt}
\begin{table}[htb]
\caption{\ti{Solving non-trivial buggy HWMCC-10 benchmarks}. \ti{Maximum number of visited states  is limited to 1,000,000 for} \Rnd~\ti{and 40,000 for} \TS}
\vspace{-10pt}
\scriptsize
\begin{center}
\begin{tabular}{|p{30pt}|p{26pt}|p{22pt}|p{23pt}|p{22pt}|p{23pt}|l|} \hline
 \Ss{number of}  & \Rnd & \multicolumn{2}{|p{45pt}|}{\TS} & \multicolumn{2}{|p{45pt}|}{\TS} & \TS\\
 \Ss{benchmarks}  & \Ss{solved} & \multicolumn{2}{|p{45pt}|}{\Ss{unrandomized}} & \multicolumn{2}{|p{51pt}|}{\Ss{randomized}} & \Ss{total}\\
\cline{3-4}\cline{5-6} 
 &   & \Ss{solved.} & \Ss{converg.}     &\Ss{solved}  & \Ss{converg.}    & \Ss{solved} \\  \hline
\Ss{236} & \tb{43}   & 35   & 94  & 59 & 8 & \tb{69} \\ \hline
\end{tabular}
\label{tbl:rand_vers_tap}
\end{center}
\end{table}

The first column of Table~\ref{tbl:rand_vers_tap} shows the number of non-trivial benchmarks used in the first experiment.
The second column gives the number of benchmarks solved by \Rnd. The parameters $max\_tries$ and $max\_length$ of \Rnd~were
set to 10,000 and 100 respectively. That is \Rnd~generated up to 10,000 counterexamples of length 100.
(So the total number of visited states was limited by 1,000,000. 
The counterexample length of 100  was large enough to solve
any benchmark solved by \TS.) For every benchmark, the time limit  for \Rnd~was set to 900 seconds. 

The next four columns show results of unrandomized and randomized versions of \TS.   For both versions,
the number of visited states (i.e. the size of \All) was limited by 40,000 and the time limit was set to 180 seconds.
For either version, we report the number of solved benchmarks and the number of benchmarks where a convergence point
was reached. (Recall that a convergence point is reached by \TS~when the set \Acts~becomes empty before a bug is found.)
 The last column gives the number of benchmarks
solved by at least one version of \TS.

The results of Table~\ref{tbl:rand_vers_tap} show that for many benchmarks the unrandomized version of \TS~  
reached a convergence point. This means that 
\TS, in its current form, needs some way to escape early convergence. In this experiment, we achieved this goal
by randomizing \TS~as described in Subsection~\ref{subsec:impl_tap_seq}. The randomized version
of \TS~solved more benchmarks and reached a convergence point only for 8 benchmarks. 
Overall,  the experiment showed that \TS~outperformed \Rnd~solving more benchmarks (69 versus 43) with much stricter limit on
the number of visited states. 

%
%
\subsection{Second Experiment: Bounded Model Checking and \TS}
\label{subsec:second_exper}
The objective of the second experiment was to show that some benchmarks solved by \TS~were 
hard for  Bounded Model Checking (BMC)~\cite{bmc}. 
In this experiment, we used a BMC tool built on top  of  the Aiger package~\cite{aiger} and Picosat~\cite{picosat}, a well-known SAT-algorithm.
In general, BMC is good at detecting shallow bugs but struggles to find deeper bugs even if these bugs are easy to detect.
This point is illustrated by results of the second experiment shown in Table~\ref{tbl:bmc_versus_tap}.  Notice that we do not claim
that the current implementation of \TS~outperforms BMC. The latter performed extremely well on shallow benchmarks
of the set we used in the first experiment  while \TS~could not solve many of them. We just
want to emphasize the promise of \TS~in finding deep bugs.

%
%
\setlength{\abovecaptionskip}{2pt}   
\setlength{\belowcaptionskip}{5pt}   
\setlength{\intextsep}{1pt}
\setlength{\floatsep}{1pt}
\begin{table}[htb]
\caption{\ti{Some benchmarks that are hard for BMC and easy for \TS}}
\vspace{-10pt}
\scriptsize
\begin{center}
\begin{tabular}{|p{49pt}|p{24pt}|p{30pt}|p{24pt}|p{30pt}|} \hline
 \Ss{benchmarks}  &  \multicolumn{2}{|p{54pt}|}{\ti{BMC}} & \multicolumn{2}{|p{54pt}|}{\TS} \\
\cline{2-3}\cline{4-5} 
  & \Ss{time (s.)} & \Ss{cex length}     &\Ss{time (s.)}  & \Ss{cex length}    \\  \hline
pdtswvroz10x6p0 & 118  & 58  & 1.2  & 88 \\ \hline
pdtswvsam6x8p0  & 116 & 48  & 7.7  &  48\\ \hline
pdtswvtma6x6p0  &95  & 57  & 0.8  & 57  \\ \hline
pdtswvtma6x4p0  & 70  & 57  & 0.9  & 57  \\ \hline
pdtswvroz8x8p0  & 65  & 48  & 1.1  & 72 \\ \hline
visbakery       & 925 & 59  & 28  & 61 \\ \hline
\end{tabular}
\label{tbl:bmc_versus_tap}
\end{center}
\end{table}

The first column of Table~\ref{tbl:bmc_versus_tap} gives benchmark names.
The next two columns show the time taken by the BMC tool to find a counterexample and the length of this counterexample.
The last two columns provide the same information for \TS.
The examples of Table~\ref{tbl:bmc_versus_tap} have the largest counterexample length among the benchmarks solved by \TS.
These are also the examples (among those solved by \TS) where the BMC tool had the longest run time. \TS~significantly
outperforms the BMC tool on these examples. Interestingly,
the first five benchmarks were also easy for \Rnd~(but \Rnd~failed to solve the 'visbakery' benchmark).

\section{Conclusions}
In this paper, we introduce \TS, a new algorithm for generation of tests for sequential circuits based on the Tests-As-Proofs (TAP) paradigm. 
\TS~forms a counterexample from encodings of proofs of local properties that are versions
of the property to be verified. The preliminary experimental results allows one to conclude that
\begin{itemize}
\item \TS~convincingly outperforms a random algorithm;
\item  \TS~significantly outperforms a BMC tool on some benchmarks with non-shallow bugs.
\end{itemize}
These results suggest that algorithms based on the TAP paradigm can be used for finding deep bugs.

Our future research will be focused in the following directions.

1) In this paper, we consider an algorithm mimicking forward model checking.
That is one generates a set of states reachable from an initial state trying to find a state violating the property  in question.
Instead, one can try to mimic a backward model checking algorithm building a set of states from which a bad state is reachable.
The objective here is to reach an initial state. Moreover, one can try to design an algorithm that combines forward and backward model
checking. Intuitively, such an algorithm can be much more effective in finding a bug because a counterexample is built from both
initial and bad states.

2) The other important direction for research is to find a better way to avoid reaching a convergence point i.e. the situation where
no new states are generated.
In this paper, we achieved this goal by randomizing the part of \TS~that performed proof encoding. This solution is not quite satisfactory
because it leads to generating too many states per time frame and hence makes it much harder for \TS~to find a deep bug. (In particular,
the benchmarks with non-shallow bugs shown in Table~\ref{tbl:bmc_versus_tap} were solved  by the unrandomized version of \TS.)

\section{Acknowledgments}
This work was supported in part by C-FAR, one of six centers of STARnet, an SRC program sponsored by MARCO and DARPA.
It was also partially funded  by NSF grant CCF-1117184. 
\bibliographystyle{plain}
\bibliography{short_sat}
\end{document}